# Synthesis and characterisation of thin-film platinum disulfide and platinum sulfide

*Conor P. Cullen[1,2], Cormac Ó Coileáin [1,2], John B. McManus[1,2], Oliver Hartwig[3], David McCloskey[4], Georg S. Duesberg[1,2,3], Niall McEvoy[1,2]*

[1] School of Chemistry, Trinity College Dublin, Dublin 2, D02 PN40, Ireland

[2] AMBER Centre, CRANN Institute, Trinity College Dublin, Dublin 2, Ireland

[3] Institute of Physics, EIT 2, Faculty of Electrical Engineering and Information Technology, Universität der Bundeswehr München, 85579 Neubiberg, Germany

[4] School of Physics, Trinity College Dublin, Dublin 2, Ireland

## Abstract

Group-10 transition metal dichalcogenides (TMDs) are rising in prominence within the highly innovative field of 2D materials. While PtS$_2$ has been investigated for potential electronic applications, due to its high charge-carrier mobility and strong layer-dependent bandgap, it has proven to be one of the more difficult TMDs to synthesise. In contrast to most TMDs, Pt has a significantly more stable monosulfide, the non-layered PtS. The existence of two stable platinum sulfides, sometimes within the same sample, has resulted in much confusion between the materials in the literature. Neither of these Pt sulfides have been thoroughly characterised as-of-yet. Here we utilise time-efficient, scalable methods to synthesise high-quality thin films of both Pt sulfides on a variety of substrates. The competing nature of the sulfides and limited thermal stability of these materials is demonstrated. We report peak-fitted X-ray photoelectron spectra, and Raman spectra using a variety of laser wavelengths, for both materials. This systematic characterisation provides a guide to differentiate between the sulfides using relatively simple methods which is essential to enable future work on these interesting materials.

## Introduction

Transition metal dichalcogenides (TMDs) are a fast growing and highly researched area in modern materials science. The numerous potential advances, made possible by their oft-quoted varied and layer-dependent properties, has led to a proportionate "gold rush" across the periodic table for suitable materials to be synthesised on the nanoscale[1-4]. This has resulted in great progress in the synthesis and understanding of TMDs[5,6]. While there is a long history to their study[7], recently the group-10 TMDs, including $PtSe_2$, have risen to a point of prominence thanks to their impressive theoretical electronic-transport capabilities[8-10]. This, in turn, has expanded the focus given to the other Pt chalcogenides (or Pt-based TMDs), $PtS_2$ and $PtTe_2$[11,12].

A consequence of the "gold rush" haste is that occasionally the due diligence in terms of accurate and refined characterisation of the TMD material is overlooked in order to focus on more forward-looking aspects such as fabrication of electrical devices or electrochemical applications. This has resulted in a lack of coherency within literature for the platinum sulfides. Indeed, for the two most common sulfides, $PtS_2$ and PtS, several publications report conflicting identification of synthesised materials[13-17]. These publications are largely reliant on Raman or X-ray photoelectron spectroscopy (XPS) for characterisation of the TMDs. Both these characterisation methods are among the most ubiquitous in 2D material characterisation, however both have their own particular limitations and can be subject to misinterpretation. Similar issues with competing sulfides have been encountered for tin sulfides where thorough comparative characterisation was needed to provide clarity[18]. Confusion in the literature has recently become a note of warning in the wider XPS community which encourages a greater focus be put upon educational material in journal publications and a widening of reviewer panels to ensure expert coverage of all aspects of a body of work[19-23].

Edmund Davy, a foundational figure in Irish chemistry[24-27], published what is to our knowledge the first paper on the synthesis of both sulfides of platinum in late 1812[28]. He was the first to recognize the utility of combing both Pt and S in an evacuated tube and heating at high temperature, he used this to synthesise PtS. A distillation setup combining an ammonium chloride of platinum and sulfur powder over mercury yielded $PtS_2$. Both materials were analysed using rudimentary characterisation methods including by taste and smell. $PtS_2$ was found to be thermally stable in anaerobic environments but to degrade quickly under ambient annealing[28]. The synthesis and characterisation of both platinum sulfides presented here can, to a certain extent, be seen as a sequel to this fundamental work with platinum sulfides.

A naturally occurring form of PtS known as cooperite was first described in 1928 by Richard A. Cooper[29], it is a significant platinum ore and therefore many of the publications concerning platinum sulfides are mineralogical in nature[30-34]. The long history of misidentification between the platinum sulfides is present here with Cooper originally proposing that the cooperite material he described had the formula $PtS_2$, cooperite was quickly re-ascribed as being PtS[35]. $PtS_2$ is not known to form naturally, indicating a thermodynamic preference for PtS.

While PtS has benefited from more attention, its study has been confined to the lens of mineralogical investigation, it has generally not been investigated or characterised heavily in the manner which modern 2D materials typically are, with only few exceptions[36,37].

Recently PtS has been observed to have several interesting properties including ultrafast saturable absorption and peculiar thickness-dependent surface states for a non-layered material[36]. PtS undergoes a pressure-induced phase change at ~3 GPa to the PdS structure $P4_2m$[38]. It has also been shown to be one of only a handful of materials which possesses negative linear compressibility, meaning the structure expands along one direction when compressed uniformly[39, 40].

$PtS_2$ has garnered recent attention in part due to its predicted properties. $PtS_2$ is the only Pt-based TMD which is semiconducting in bulk[41]. It has a strong layer-dependent bandgap variation with a change of the indirect bandgap from 0.25-1.6 eV when going from bulk to monolayer[41]. $PtS_2$ is predicted to have strongly-bound excitons in the monolayer form[42] and a calculated monolayer electron mobility of up to ~3900 $cm^2$ $V^{-1}$ $s^{-1}$ [10], with FET devices delivering room-temperature electron mobility of 62.5 $cm^2$ $V^{-1}$ $s^{-1}$ and a $10^6$ on/off ratio[43].

The majority of recent $PtS_2$ experimental studies have used material synthesised by chemical vapour transport (CVT) methods, this can then be exfoliated to give individual 2D crystals. While this can result in high-quality material, this is not guaranteed, with PtS contamination being common[14, 43-46]. The main disadvantages of using CVT-style methods to synthesise and isolate 2D materials are that they are time consuming, labour-intensive and inherently unscalable.

While $PtS_2$ and PtS have both been synthesised via chemical vapour deposition (CVD) methods[36, 47, 48]. Robust metrics and characterisation to confirm the quality and purity of these materials are noticeably absent.

Although the Raman spectrum of few-layer $PtS_2$ has been studied in several works, it is an area with many unexplored aspects[43-46]. The catalogue of PtS Raman is poorer, with only a few works providing any insight into the features of the complex spectrum[31, 36], with some published PtS Raman spectra incorrectly claiming to be $PtS_2$ adding to the confusion.

High-resolution XPS spectra of these materials are provided in several pieces of literature[14, 17, 49-51], unfortunately there exists little to no peak-fitted deconvolution of these materials that holds up to best-practice scrutiny.

This work describes a relatively simple methodology to synthesise high-purity thin films of both PtS and $PtS_2$ using systems and equipment common to many TMD synthesis labs. Both platinum sulfides are rigorously characterised using Raman and XPS to better the understanding of the important aspects of their spectra. The stability of these films and the ability to convert $PtS_2$ into PtS are also investigated.

**Results and discussion**

### Synthesis of PtS

Polycrystalline films of PtS were synthesised using a procedure previously described for other TMDs[52]. A Pt metal layer is deposited onto a substrate, typically $SiO_2$/Si, by sputtering and then placed into a two-zone quartz-tube furnace with sulfur powder in the secondary zone as shown in Fig.1(b). Annealing the Pt film at 500 °C under forming gas (90% Ar/10% $H_2$) at a pressure of ~1 mbar for 1 hour generated uniform films of PtS.

## Synthesis of PtS$_2$

Polycrystalline films of PtS$_2$ were synthesised using a modified design inspired by the work of Wu et al.[48]. Using the same two-zone quartz tube furnace as for PtS, the Pt metal film was placed inside an internal open-ended quartz tube in the main heating zone. The opening of this internal tube, at the edge of the hot-zone, was loaded with sulfur powder as a second local sulfur source and the opening placed against the direction of flow, as shown in Fig.1(a). The effect of this is to trap a high partial pressure of sulfur in the internal tube. Annealing at 500 °C under Ar/forming flow and a pressure of ~200 mbar for 1 hour generated uniformly converted films of PtS$_2$.

More details of synthesis procedure for both materials can be found in the methods section. An advantage of these processes is the ability to synthesise films on a variety of substrates. PtS$_2$ and PtS films were synthesised on SiO$_2$/Si, pyrolytic carbon (PyC), and quartz substrates as shown in the supplementary information, Fig.S1. These are applications-oriented substrates frequently used for electronics, electrochemistry, and optics respectively.

The effects of synthesis temperature and initial Pt film thickness on the resulting TMD film were investigated for both materials. For simplicity, to differentiate between films they will be referred to by their initial Pt metal film thickness. The expansion factor when sulfurizing Pt films to PtS has been shown to vary depending on starting thickness[36]. It has previously been shown that a similar process for PtSe$_2$ results in an approximate quadrupling of film thickness after conversion.[53]

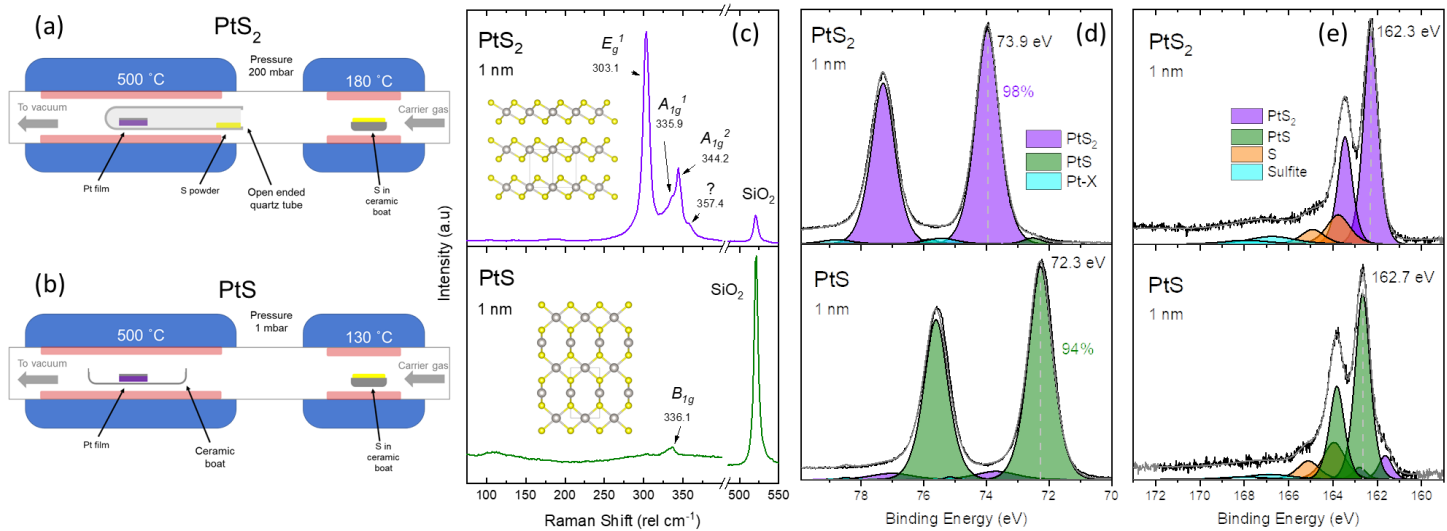

**Figure 1.** Schematic diagram for the synthesis arrangement for PtS$_2$ films **(a)** and PtS films **(b)**. **(c)** Raman spectra of PtS$_2$ and PtS films synthesized from a 1 nm Pt film, with an insert showing their atomic structures (yellow balls represent sulfur atoms, silver balls represent platinum atoms). XPS of the Pt 4f **(d)** and S 2p **(e)** core-level regions for PtS$_2$ and PtS films.

**Raman spectroscopy of PtS and PtS$_2$**

Raman spectra of 1 nm PtS$_2$ and 1 nm PtS, taken using a 532 nm excitation wavelength, are shown in Fig.1(c). PtS$_2$ has an octahedral (1T) geometry with a hexagonal lattice, and a $D_{3d}$ point group[46], see supplementary information Fig.S2(a). Therefore the three primary components to the Raman spectrum of few-layer PtS$_2$ are the out-of-plane $E_g^1$ mode at ~303 rel cm$^{-1}$, and the two in-plane modes $A_{1g}^1$ and $A_{1g}^2$ at ~336 and 344 rel cm$^{-1}$ respectively[46]. These components are clearly visible in Fig.1(c) alongside an unidentified shoulder at ~357 rel cm$^{-1}$. PtS$_2$ displays high intensity Raman signals compared to the underlying SiO$_2$.

The position of these peaks aligns with reports for ~2-3-layer PtS$_2$[46]. The full width at half maximum (FWHM) of the peaks in this spectrum are consistent with those reported for similar growth methods but are significantly broader than those reported for individual mechanically-exfoliated flakes[44, 46, 48]. While the FWHM of Raman peaks is affected by laser power and the spectral grating, the broadening here is comparable to what has been reported for other TMDs when comparing polycrystalline films to large single crystals from CVD[54, 55].

To provide a thorough database of characterisation, the Raman spectra of a 5 nm PtS$_2$ film using three excitation wavelengths (532, 633, and 405 nm) is given in the supplementary information, Fig.S3. No new modes or significant changes in intensity ratios between the peaks are observed for the different wavelengths.

PtS is a non-layered material with a tetragonal crystal structure, see Fig.S2(b), and either a P4$_2$/*mmc*[31, 34] or P4$_2$/*nmn* space group[39]. The Raman spectrum of the 1 nm PtS film in Fig.1(c) shows a single low intensity peak at ~336 rel cm$^{-1}$. This is consistent with other reports of PtS and is attributed to the Raman-active $B_{1g}$ vibrational mode[36]. The 100-150 rel cm$^{-1}$ region shows a broad background for very thin PtS films, which develops into distinct peaks for thicker films. Fig.S3 in the supplementary information shows the Raman spectrum of a 5 nm PtS film acquired with the three different excitation wavelengths. The 532 nm and the 633 nm spectra show several clearly resolved peaks. These spectra are distinct from each other with the 532 nm spectrum being dominated by the $B_{1g}$ peak at ~335 rel cm$^{-1}$ with many lower intensity peaks across the spectrum. The 633 nm spectrum has four sharp peaks at approximately 115, 335, 377, and 475 rel cm$^{-1}$. These various additional peaks in the 532 nm spectrum and the 633 nm Raman of PtS were previously unreported. This disparity in Raman spectra between the excitation energies is likely a result of resonance effects, the additional Raman modes are currently unassigned in literature and should form the basis for further study.

**XPS of PtS and PtS$_2$**

XPS was used to characterise both of these films as shown in Fig.1(d, e). XPS has extreme utility in the characterisation of nanoscale materials as it has very high surface sensitivity, can differentiate between chemical states, and allows calculation of stoichiometry amongst other measures. The drawback when utilising XPS is the non-trivial nature of interpreting core-level spectra. The data can be complex, leaving significant room for misinterpretation through fitting errors and these uncertainties can be difficult to convey fairly when reporting data. Numerous reports have been published recently in an effort to improve the overall quality of XPS reported in the literature[20, 22]. Due to the complicated nature of the XPS this section provides a resource to help deconvolute the spectra of platinum-sulfur compounds.

The spectral regions of interest for XPS of platinum sulfides are the Pt 4f and S 2p core-levels. The XPS spectra of the Pt 4f region of 1 nm $PtS_2$ and PtS films is shown in Fig.1(d). The spectrum for $PtS_2$ is dominated by a doublet pair corresponding to the Pt(IV) state for $PtS_2$, with the $Pt\ 4f_{7/2}$ component at 73.9 eV. This accounts for 98% of the Pt atoms measured, indicating the high purity of material obtained. A low-binding energy shoulder with $Pt\ 4f_{7/2}$ at ~72.5 eV is attributed to a Pt(II) state from PtS. The PtS component in the spectrum could potentially be misidentified and assigned as Pt metal due to the relatively small difference in expected binding energy position. While basing peak assignments on literature values of energy position alone has issues[21], the symmetric peak shape of the PtS component makes a Pt metal state very unlikely. Low levels of PtS were found to be almost ever present in all of the $PtS_2$ films synthesised here. A third component with $Pt\ 4f_{7/2}$ at ~75.4 eV is fitted, assignment of this low intensity and broad doublet is difficult and it is provisionally assigned as a Pt-oxide due to the binding energy of the peaks, this aligns with previous interpretations for $PtSe_2$[56].

Similarly the Pt 4f spectrum for a PtS film has a doublet with $Pt\ 4f_{7/2}$ at 72.3 eV, this is attributed to Pt(II) from PtS. The PtS component accounts for 94% of the measured Pt atoms. A secondary doublet with $Pt\ 4f_{7/2}$ at ~73.8 eV is from Pt(IV) in the form of $PtS_2$.

The S 2p core-level region is significantly more complex and has been frequently oversimplified in the literature[14, 17, 36, 51]. Similar line-shapes to those shown in this work have been observed but not fitted for the platinum sulfides previously[49, 50]. The S 2p regions for both platinum sulfides are fitted with four doublets from four chemical states of sulfur. $PtS_2$ and PtS components are on the low binding energy side with their $S\ 2p_{3/2}$ peaks at 162.3 and 162.7 eV respectively. Elemental sulfur was found on both samples at ~163.8 eV. The fourth component was a consistently low intensity broad peak with $S\ 2p_{3/2}$ at ~166.5 eV. While similar line-shapes have been reported for platinum sulfides previously, there has been little discussion of their origin[15]. We believe the most likely chemical state for this broad component is a S-O species, likely sulfite based on binding energy position[57]. Oxidised sulfur is not commonly observed in XPS for other sulfur TMDs synthesised through similar methods[52], this implies that Pt may play a role in the oxidation,[58, 59] or that this broad component has a different origin.

Composite materials (consisting of both PtS and $PtS_2$) can be readily obtained. A film of both PtS and $PtS_2$ in approximately equal amounts can be synthesised by lowering the synthesis temperature to 400 °C with a reduced amount of sulfur in the internal tube. The Raman and XPS spectra for this are shown in the supplementary information, Fig.S4. This allows us to more clearly illustrate the differences in XPS peak positions from the different chemical states.

The overlapping nature of many of the components in both the Pt 4f and the S 2p core-levels increases the possibility for error in each fitting, small changes in relative peak areas can result in large changes in the calculated stoichiometry of the material. In an effort in minimise this error, we averaged stoichiometry values over several samples (see supplementary information, Fig.S5) giving S:Pt ratios of 1.87±0.06 for $PtS_2$ and 0.97±0.09 for PtS. These values are very close to the ideal values, the chalcogen deficiency for $PtS_2$ potentially indicates that sulfur vacancies are a common defect similar to other TMDs[60].

## Thickness dependence of spectral features

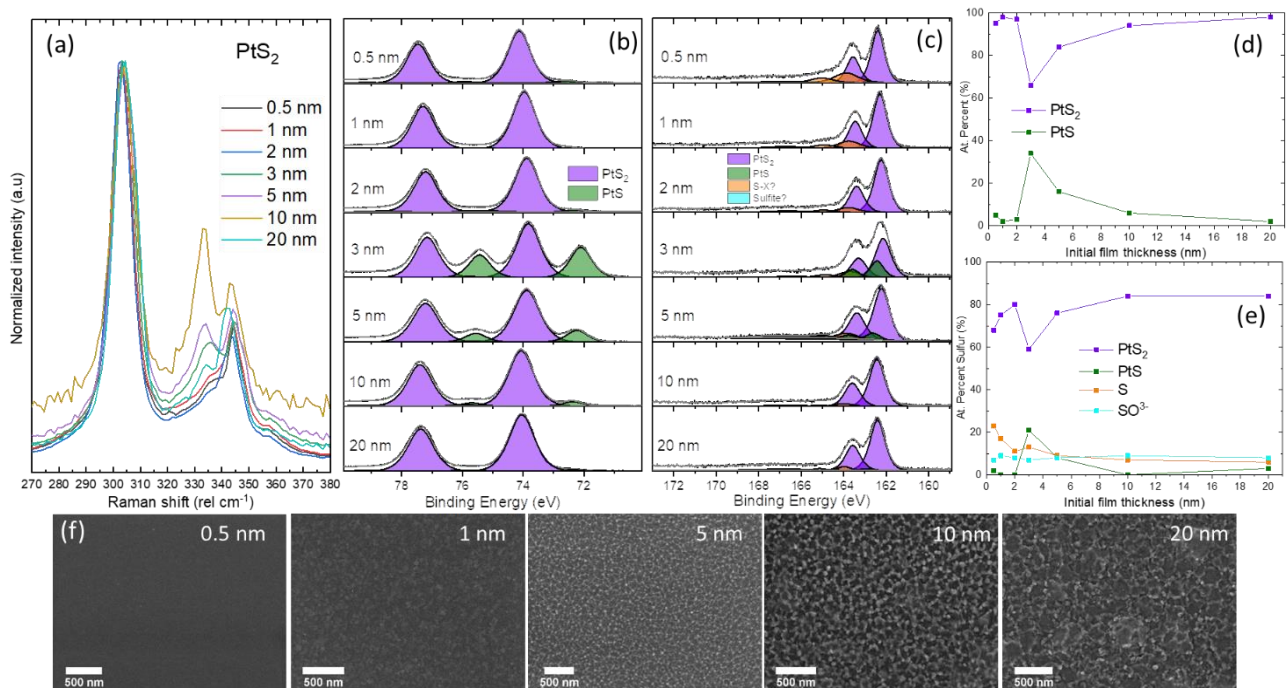

**Figure 2. (a)** Raman spectra of PtS$_2$ films of different thickness, normalized to the $E_g^1$ peak intensity. XPS of the Pt 4f **(b)** and S 2p **(c)** core-level regions for PtS$_2$ films of different thickness. Chemical composition from XPS of the Pt 4f **(d)** and S 2p **(e)** XPS regions. **(f)** SEM images of PtS$_2$ films of different thickness.

The thickness dependence of this synthesis procedure was investigated for PtS$_2$. Raman, XPS spectra, chemical composition data from XPS, and scanning electron microscopy (SEM) images for 7 thicknesses of PtS$_2$ films are shown in Fig.2. The Raman spectra for PtS$_2$ films with increasing Pt film thickness Fig.2(a) show only minor changes for the $E_g^1$ and $A_{1g}^2$ Raman modes. It is expected from the literature that with increasing thickness the $A_{1g}^2$ and $A_{1g}^1$ modes gradually merge, with bulk PtS$_2$ having only two Raman peaks for the $E_g^1$ and the $A_{1g}$ [41]. We see an inconsistent trend here due to uniformity issues for thicker films, but it is noteworthy that for films of up to 20 nm starting Pt thickness we can still differentiate between the two $A_{1g}$ modes indicating we have not reached bulk-like behaviour. The increasing prominence of what appears to be the $A_{1g}^1$ at ~334 rel cm$^{-1}$ is attributed to a convolution of the $A_{1g}^1$ from PtS$_2$ and the $B_{1g}$ mode from PtS contamination, this overlap of the Raman modes contributes to the possible errors in identification of the Pt sulfides.

The weak adhesion of Pt films on SiO$_2$ is a well-known issue[61, 62], this was found to impact this system with films of >5 nm starting Pt thickness being prone to bubbling and delaminating from the SiO$_2$. This could potentially be remedied by use of an adhesion layer between the Pt and the substrate, although common adhesion materials such as Ti may be problematic as they have the potential to sulfurize during synthesis or to leech chalcogen from the platinum sulfide films.[56]

XPS spectra for the Pt 4f and S 2p core-level regions PtS$_2$ films of various thicknesses are shown in Fig.2(b,c) with the chemical state composition amassed from XPS for each core-level shown in Fig.2(d,e).

The synthesis of PtS$_2$ was successful at almost all thicknesses examined, with the XPS core-levels of the films showing a consistently high level of PtS$_2$. The exception were the films synthesised from a starting Pt film of 3 nm. These were consistently observed to poorly convert to PtS$_2$ with the sample shown here only yielding a 65:35 ratio of PtS$_2$:PtS. The cause of this peculiarity is not currently understood and merits further study, PtSe$_2$ thin films have been shown to change orientation with increasing metal film thickness,[53] a similar effect could potentially play a role in this system also.

A combination of the shallow measurement depth of XPS and the delamination/bubbling of the PtS$_2$ films, discussed below, results in potentially misleading results from XPS if taken alone. This illustrates why a combination of measurement techniques, as implemented here with Raman spectroscopy, is the minimum requirement to gain a clear understanding of the entire film's properties and composition.

For films with thicknesses below the limit of film delamination from the substrate (~5 nm), the large-area consistency of the films can be seen in Fig.S6–S8 in the supplementary information, in which optical microscopy images and scanning Raman spectroscopy maps of the surface are shown.

SEM images of the PtS$_2$ films are shown in Fig.2(f), these images make the highly polycrystalline nature of the films clear and look similar to other TMD films synthesised through related methods.[63] Contrast increases with thickness of the films, individual crystallites on the order of 50-100 nm are visible across the PtS$_2$ films, with average crystallite size generally increasing with film thickness. While minor changes were observed in film morphology for starting thicknesses less than 5 nm (see also supplementary information, Fig.S9.), this changes substantially over 5 nm, with a combination of effects including the lack of adhesion, expansion of the films, and strain during synthesis resulting in bubbling and delamination for thicker films. Furthermore, for the films with starting thicknesses less than 5 nm, which exhibited minor changes in SEM images, insight into the film morphology is provided using atomic force microscopy (AFM), as shown in the supplementary information Fig.S10–S12.

Variations as a result of sample position inside the internal quartz tube were investigated with no significant variance observed in XPS or Raman of the films indicating the robust nature of this procedure. (see supplementary information, Fig.S13.)

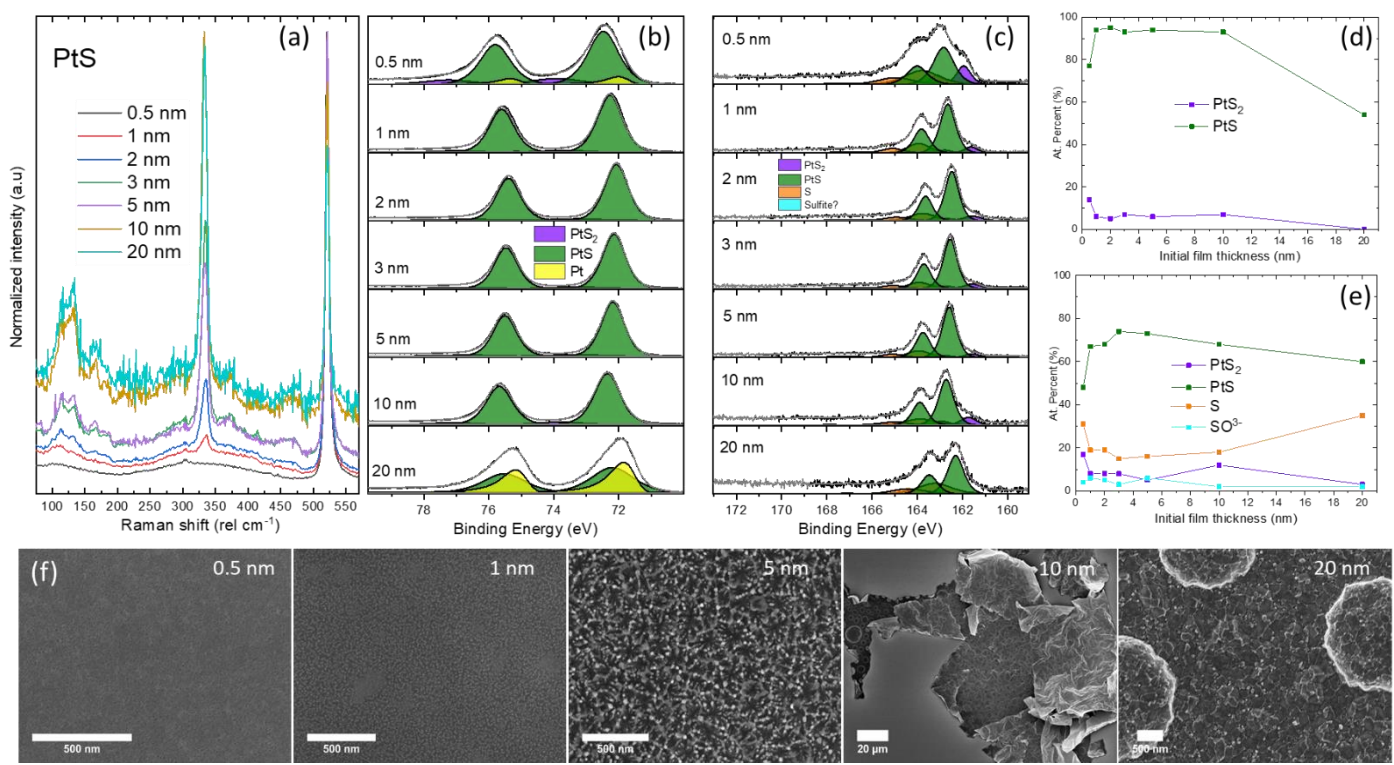

**Figure 3. (a)** Raman spectra of PtS films of different thickness. XPS of the Pt 4f **(b)** and S 2p **(c)** core-level regions for PtS films of different thickness. Composition of the Pt 4f **(d)** and S 2p **(e)** XPS regions. **(f)** SEM images of PtS films of different thickness.

Similarly, the thickness dependence of PtS films was investigated. The Raman spectra of PtS films are shown in Fig.3(a) to have a strong thickness dependence. Very thin films are seen to show no discernible Raman peaks but have two very broad regions centred at ~330 and ~100 rel cm$^{-1}$. As the thickness of the film increases more pronounced modes appear, first with the $B_{1g}$ at ~335 rel cm$^{-1}$ followed by a number of smaller peaks across the spectrum. Two heavily overlapping sharp peaks are seen to develop at ~115 and 134 rel cm$^{-1}$. This is the first report of these Raman modes and their thickness dependence for PtS, they are currently unassigned.

XPS analysis of the PtS films in Fig.3 (b,c) demonstrates the advantage of complementary characterisation with Raman and XPS. While the Raman spectrum from the 0.5 nm film does not show any characteristic PtS Raman modes, in contrast XPS indicates the film is predominantly PtS. Interestingly, there is a small amount of Pt metal present in the 0.5 nm sample indicating that either the film was not fully converted initially or that the film was more prone to degradation upon exposure to ambient conditions. 1 nm–5 nm PtS films gave XPS spectra with high levels of PtS and no Pt metal present, PtS films in this thickness bracket generally showed very similar line shapes, indicating high uniformity. The large-area homogeneity of these films can be seen in the supplementary information, Fig.S6–S7, in which optical microscopy images and scanning Raman spectroscopy maps of the surface are shown. Significant amounts of Pt metal were present for the 20 nm film, this is most likely a result of incomplete conversion for thicker films combined with extensive delamination exposing the lower, unconverted layers. Pt thicknesses >5 nm were very prone to delamination, this can be exacerbated by higher temperatures, increasing the likelihood and degree of delamination.

SEM of the PtS films in Fig.3(f) show relatively uniform films with minimal contrast for an initial 1 nm Pt film. Larger crystallites and an uneven surface appear at ~5 nm thickness, with significant bubbling, delamination and folding occurring for thicker films. This can be seen clearly in the SEM images for the 10 and 20 nm PtS films. Additional surface morphology information by means of AFM is provided in the supplementary information, Fig.S10–S12.

## Conversion of PtS$_2$ to PtS: Annealing and PtS$_2$ stability.

The thermally driven transformation of PtS$_2$ to PtS is analysed at several temperatures using complementary Raman and XPS. Fig. 4 shows the XPS and Raman data for four 1 nm PtS$_2$ films with post-growth annealing in a quartz tube at 1 mbar under Ar flow for 30 minutes.

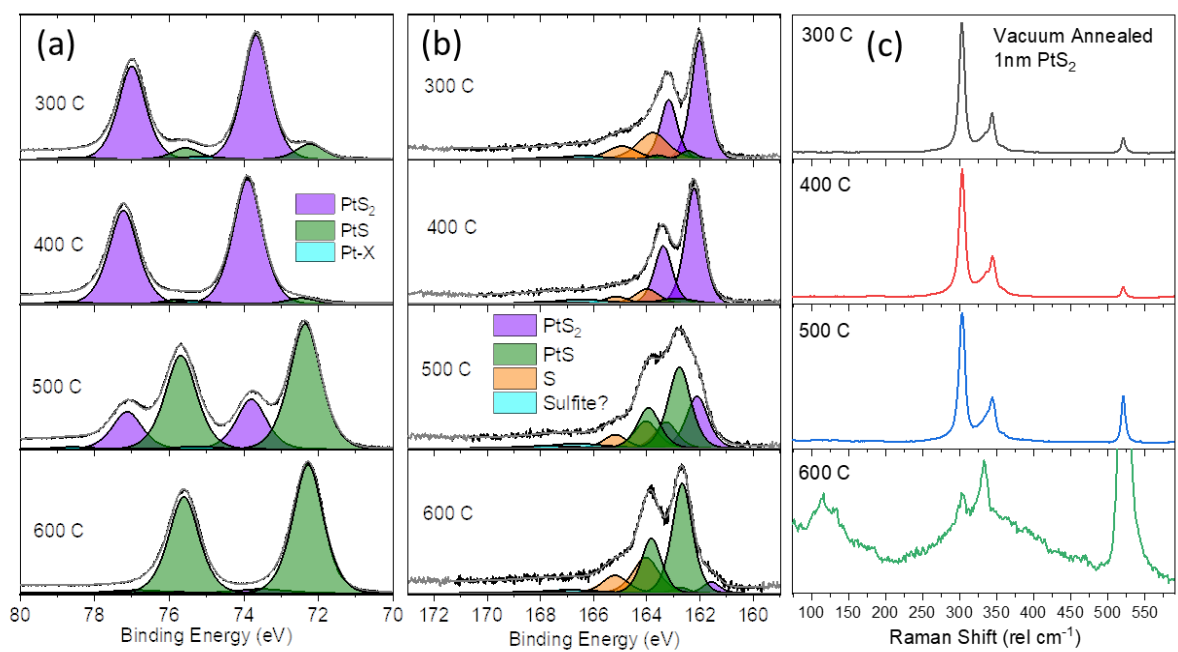

**Figure 4.** XPS of the Pt 4f **(a)** and S 2p **(b)** core-level regions for 1 nm PtS$_2$ films after annealing at various temperatures. **(a)** Raman spectra of 1 nm PtS$_2$ films after annealing at various temperatures

Little change occurs when compared to the pristine material (Fig.2), including no reduction of the PtS$_2$ to PtS for the samples annealed at either 300 or 400 °C, with variations in their PtS content attributed to normal sample-to-sample variation in the starting PtS$_2$ films.

500 °C annealing results in majority conversion of the PtS$_2$ film into PtS with a 27:73 ratio of PtS$_2$:PtS. Despite the large majority of PtS in the film, the Raman spectrum is only slightly altered from that of pure PtS$_2$, this is a potentially major factor in mischaracterisation of Pt sulfide films. The disparity is a result of the Raman signal for PtS$_2$ being significantly more intense than that of PtS, illustrating that Raman alone is not a reliable technique for characterisation of these materials.

This may explain the discrepancy with the results of Wu *et al*. who reported the reversible conversion of the Pt sulfides: Annealing at 525 °C converting $PtS_2$ into PtS, while annealing at 450 °C in a sulfur vapour converts PtS into $PtS_2$, both were characterised using only Raman spectroscopy[48].

Annealing at 600 °C yields an XPS spectrum in Fig. 4(a) with 95% PtS with a corresponding pronounced change in the Raman spectrum Fig. 4(c), the characteristic PtS Raman mode at ~336 rel cm$^{-1}$ and the broad modes between 100-150 rel cm$^{-1}$ are all visible. A remnant of $PtS_2$ Raman signal is still visible with the $E_g^1$ at ~300 rel cm$^{-1}$, further indicating the imbalance in their Raman signal intensity. Annealing of 2 and 3 nm $PtS_2$ films at 600 °C similarly yields full conversion of the $PtS_2$ Raman signal to give PtS like Raman signal (see supplementary information, Fig.S14).

Similar conversion of $PtSe_2$ to the non-layered PtSe in UHV conditions above 500 °C has been demonstrated previously, but unlike the conversion of $PtS_2$ to PtS the monochalcogenide PtSe is less stable than its dichalcogenide counterpart[64].

While stability at 400 °C is significant for a few-layer material this should be seen as an upper limit for thermal stability of $PtS_2$ in an anaerobic environment. This thermal stability is a potentially important factor when considering the compatibility of materials in device fabrication.

**Conclusions**

In summary, we have synthesised polycrystalline thin films of both $PtS_2$ and PtS on a variety of substrates using direct sulfurization methods in a quartz-tube furnace in a time efficient process. We have reported fitted XPS spectra revealing the components responsible for the complex peak structure of the platinum sulfides. Raman spectra of the platinum sulfides were examined over several thicknesses using three excitation laser lines, we observed substantial variation in the number and proportion of Raman modes in PtS using different excitation energies. While they are frequently confused in the literature, the complementary characterisation presented here allows the two materials to be unambiguously distinguished from one another. This report provides needed clarity by thorough comparative characterisation of both platinum sulfides and will hopefully act as a reference to enable future studies with these emerging materials. Thermal processability limits for thin films of $PtS_2$ were examined and shown to result in gradual conversion to PtS for anaerobic annealing over 400 °C.

**Experimental**

**Film deposition**

Pt films were deposited on $Si/SiO_2$ substrates with controlled thickness using a Gatan precision etching and coating system (PECS 682). Thickness and deposition rate were closely monitored using a quartz-crystal monitor.

**Synthesis of PtS (Schematic diagram in Fig.1)**

Substrates were placed in the primary heating zone of a two-zone quartz-tube furnace with a crucible of sulfur powder placed in the secondary heating area. The substrates were heated to 500 °C under a forming gas environment (90% Ar/10% $H_2$, 150 sccm) with continuous vacuum pumping establishing a ~1 mbar atmosphere. When the substrates reached 500 °C the sulfur powder was heated to 130 °C for 1 hour. After this, the forming gas was changed to 100% Ar and the furnace was then turned-off, opened, and air cooled to room temperature before samples were retrieved.

**Synthesis of $PtS_2$ (Schematic diagram in Fig.1)**

The same system as described for PtS growth was used for $PtS_2$ synthesis. The changes were as follows. Substrates were placed into a 17 mm diameter (2 mm wall) quartz tube with one open end. Sulfur powder was placed at the entrance of this tube. This tube was then placed into the primary heating zone of the quartz tube furnace with the open side facing into the direction of flow and placed with the opening near the edge of the main heating zone. The substrates were heated to 500 °C under the same forming gas flow as previously. Vacuum pumping was manually throttled to maintain a pressure of ~200 mbar throughout the synthesis. When the substrates reached 500 °C the sulfur powder was heated to 180 °C and maintained for 1 hour. After this, the forming gas was then changed to 100% Ar and the furnace was then turned-off, opened, and air cooled to room temperature before samples were retrieved.

**Characterisation**

**Raman**

Raman spectroscopic analysis was performed at 100x magnification using WITec Alpha 300 R confocal Raman microscopes with the majority using a 532 nm excitation source at a power of <200 µW and a spectral grating of 1800 lines/mm. 633 nm and 405 nm excitation energy spectra were acquired using the same parameters. All spectra were gathered by averaging measurements over >10 discrete point spectra along the surface.

**XPS**

XPS spectra were recorded using a PHI VersaProbe III instrument equipped with a micro-focused, monochromatic Al Kα source (1486.6 eV) and dual-beam charge neutralization. Core-level spectra were taken using a pass energy of 26 eV. Spectra were processed using CasaXPS software. Spectra were charge corrected using the C 1s binding energy value of 248.8 eV, this method is imperfect but is the currently accepted standard[21]. After subtracting a Shirley type background, core-level spectra were fitted with Gaussian-Lorentzian (40:60) line shapes and a Doniach–Sunjic line shape for metallic Pt.

Stoichiometry calculations were made by comparing the relative areas of the relevant components after accounting for their relative sensitivity factors.

**SEM**

Images were acquired using a Karl Zeiss Supra microscope operating with a 3 kV accelerating voltage, 30 µm aperture and a working distance of ~3-4mm.

**Acknowledgements**

This work was supported by Science Foundation Ireland (SFI) through (15/IA/3131, 12/RC/2278_P2, 15/SIRG/3329). The SEM imaging for this project was carried out at the Advanced Microscopy Laboratory (AML), Trinity College Dublin, Ireland. The AML is an SFI supported imaging and analysis centre, part of the CRANN Institute and affiliated to the AMBER centre. G.S.D and O.H. acknowledge the European Commission under the project Queformal [829035] and the German Ministry of Education and Research (BMBF) under the projects ACDC [13N15100] and NobleNEMS [16ES1121].

**References**

1. V. Nicolosi, M. Chhowalla, M. G. Kanatzidis, M. S. Strano and J. N. Coleman, *Science*, 2013, **340**, 1226419-1226419.
2. P. Miró, M. Audiffred and T. Heine, *Chem. Soc. Rev.*, 2014, **43**, 6537-6554.
3. E. C. Ahn, *npj 2D Materials and Applications*, 2020, **4**, 17.
4. Q. H. Wang, K. Kalantar-Zadeh, A. Kis, J. N. Coleman and M. S. Strano, *Nat. Nanotechnol.*, 2012, **7**, 699-712.
5. M. Samadi, N. Sarikhani, M. Zirak, H. Zhang, H.-L. Zhang and A. Z. Moshfegh, *Nanoscale Horiz.*, 2018, **3**, 90-204.
6. W. Choi, N. Choudhary, G. H. Han, J. Park, D. Akinwande and Y. H. Lee, *Mater. Today*, 2017, **20**, 116-130.
7. F. Gr0nvold, H. Haraldsen and A. Kjekshus, *Acta Chem. Scand.*, 1960, **14**.
8. C. Yim, K. Lee, N. McEvoy, M. O'Brien, S. Riazimehr, N. C. Berner, C. P. Cullen, J. Kotakoski, J. C. Meyer, M. C. Lemme and G. S. Duesberg, *ACS Nano*, 2016, **10**, 9550-9558.
9. M. O'Brien, N. McEvoy, C. Motta, J.-Y. Zheng, N. C. Berner, J. Kotakoski, K. Elibol, T. J. Pennycook, J. C. Meyer, C. Yim, M. Abid, T. Hallam, J. F. Donegan, S. Sanvito and G. S. Duesberg, *2D Mater.*, 2016, **3**, 021004.
10. Z. Huang, W. Zhang and W. Zhang, *Materials*, 2016, **9**.
11. J. B. McManus, D. Horvath, M. Browne, C. P. Cullen, G. Cunningham, T. Hallam, K. Zhussupbekov, D. Mullarkey, C. Ó Coileáin, I. V. Shvets, M. Pumera, G. S. Duesberg and N. McEvoy, *Nanotechnology*, 2020, **31**, 375601.
12. E. Chen, W. Xu, J. Chen and J. H. Warner, *Materials Today Advances*, 2020, **7**, 100076.
13. D. Zhao, S. Xie, Y. Wang, H. Zhu, L. Chen, Q. Sun and D. W. Zhang, *AIP Adv.*, 2019, **9**, 025225.
14. Y. Wang, K. Szokolova, M. Z. M. Nasir, Z. Sofer and M. Pumera, *Chem. Eur. J.*, 2019, **25**, 7330-7338.
15. X. Wang, H. Long, W. Qarony, C. Y. Tang, H. Yuan and Y. H. Tsang, *J. Lumin.*, 2019, **211**, 227-232.
16. X. Wang, P. K. Cheng, C. Y. Tang, H. Long, H. Yuan, L. Zeng, S. Ma, W. Qarony and Y. H. Tsang, *Opt. Express*, 2018, **26**, 13055-13060.
17. X. Chia, A. Adriano, P. Lazar, Z. Sofer, J. Luxa and M. Pumera, *Adv. Funct. Mater.*, 2016, **26**, 4306-4318.

18. H. Choi, J. Lee, S. Shin, J. Lee, S. Lee, H. Park, S. Kwon, N. Lee, M. Bang, S.-B. Lee and H. Jeon, *Nanotechnology*, 2018, **29**, 215201.
19. M. R. Linford, V. S. Smentkowski, J. T. Grant, C. R. Brundle, P. M. A. Sherwood, M. C. Biesinger, J. Terry, K. Artyushkova, A. Herrera-Gómez, S. Tougaard, W. Skinner, J.-J. Pireaux, C. F. McConville, C. D. Easton, T. R. Gengenbach, G. H. Major, P. Dietrich, A. Thissen, M. Engelhard, C. J. Powell, K. J. Gaskell and D. R. Baer, *Microsc. Microanal.*, 2020, **26**, 1-2.
20. P. M. A. Sherwood, *Surf. Interface Anal.*, 2019, **51**, 589-610.
21. G. Greczynski and L. Hultman, *Prog. Mater Sci.*, 2020, **107**, 100591.
22. D. R. Baer, K. Artyushkova, C. Richard Brundle, J. E. Castle, M. H. Engelhard, K. J. Gaskell, J. T. Grant, R. T. Haasch, M. R. Linford, C. J. Powell, A. G. Shard, P. M. A. Sherwood and V. S. Smentkowski, *J. Vac. Sci. Technol. A*, 2019, **37**, 031401.
23. A. G. Shard, *J. Vac. Sci. Technol. A*, 2020, **38**, 041201.
24. H. Robert, *Oxford Dictionary of National Biography*, 1888, DOI: 10.1093/odnb/9780192683120.013.7311.
25. F. L. Christopher, *Oxford Dictionary of National Biography*, 2004, DOI: 10.1093/ref:odnb/7311.
26. J. Russell, *J. Chem. Educ.*, 1953, **30**, 302.
27. J. Wisniak, *Educación Química*, 2020, **31**, 144-155.
28. E. Davy, *The Philosophical Magazine*, 1812, **40**, 27-39.
29. R. A. Cooper, *Journ. Metall Mining Soc. South Africa*, 1928, **28**, 281.
30. S. M. C. Verryn and R. K. W. Merkle, *Can. Mineral.*, 2002, **40**, 571-584.
31. R. Pikl, D. De Waal, R. K. W. Merkle and S. M. C. Verryn, *Appl. Spectrosc.*, 1999, **53**, 927-930.
32. F. Melcher, T. Oberthür and J. Lodziak, *Can. Mineral.*, 2005, **43**, 1711-1734.
33. R. K. W. Merkle and S. M. C. Verryn, *Mineralium Deposita*, 2003, **38**, 381-388.
34. V. I. Rozhdestvina, A. A. Udovenko, S. V. Rubanov and N. V. Mudrovskaya, *Crystallogr. Rep.*, 2016, **61**, 193-202.
35. F. A. Bannister, *Mineralogical Magazine and Journal of the Mineralogical Society*, 1932, **23**, 188-206.
36. J. Huang, N. Dong, N. McEvoy, L. Wang, C. Ó. Coileáin, H. Wang, C. P. Cullen, C. Chen, S. Zhang, L. Zhang and J. Wang, *ACS Nano*, 2019, **13**, 13390-13402.
37. K. Handler, L. C. O'Brien and J. J. O'Brien, *Journal of Molecular Spectroscopy*, 2010, **263**, 78-81.
38. R. Collins, R. Kaner, P. Russo, A. Wold and D. Avignant, *Inorg. Chem.*, 1979, **18**, 727-729.
39. A. Marmier, P. S. Ntoahae, P. E. Ngoepe, D. G. Pettifor and S. C. Parker, *Phys. Rev. B Condens. Matter*, 2010, **81**, 172102.
40. A. B. Cairns and A. L. Goodwin, *Phys. Chem. Chem. Phys.*, 2015, **17**, 20449-20465.
41. R. A. B. Villaos, C. P. Crisostomo, Z.-Q. Huang, S.-M. Huang, A. A. B. Padama, M. A. Albao, H. Lin and F.-C. Chuang, *npj 2D Materials and Applications*, 2019, **3**, 1-8.
42. M. Sajjad, N. Singh and U. Schwingenschlögl, *Appl. Phys. Lett.*, 2018, **112**, 043101.
43. Z. Wang, P. Wang, F. Wang, J. Ye, T. He, F. Wu, M. Peng, P. Wu, Y. Chen, F. Zhong, R. Xie, Z. Cui, L. Shen, Q. Zhang, L. Gu, M. Luo, Y. Wang, H. Chen, P. Zhou, A. Pan, X. Zhou, L. Zhang and W. Hu, *Adv. Funct. Mater.*, 2019, **30**, 1907945.
44. L. Pi, L. Li, X. Hu, S. Zhou, H. Li and T. Zhai, *Nanotechnology*, 2018, **29**, 505709.
45. Y.-F. Yuan, Z.-T. Zhang, W.-K. Wang, Y.-H. Zhou, X.-L. Chen, C. An, R.-R. Zhang, Y. Zhou, C.-C. Gu, L. Li, X.-J. Li and Z.-R. Yang, *Chin. Physics B*, 2018, **27**, 066201.
46. Y. Zhao, J. Qiao, P. Yu, Z. Hu, Z. Lin, S. P. Lau, Z. Liu, W. Ji and Y. Chai, *Adv. Mater.*, 2016, **28**, 2399-2407.

47. W. Biltz and R. Juza, *Zeitschrift für anorganische und allgemeine Chemie*, 1930, **190**, 161-177.
48. H. Xu, H.-P. Huang, H. Fei, J. Feng, H.-R. Fuh, J. Cho, M. Choi, Y. Chen, L. Zhang, D. Chen, D. Zhang, C. Ó. Coileáin, X. Han, C.-R. Chang and H.-C. Wu, *ACS Applied Materials & Interfaces*, 2019, **11**, 8202-8209.
49. J. Dembowski, L. Marosi and M. Essig, *Surf. Sci. Spectra*, 1993, **2**, 104-108.
50. J. Dembowski, L. Marosi and M. Essig, *Surf. Sci. Spectra*, 1993, **2**, 133-137.
51. N. Rohaizad, C. C. Mayorga-Martinez, Z. Sofer, R. D. Webster and M. Pumera, *Applied Materials Today*, 2020, **19**, 100606.
52. R. Gatensby, T. Hallam, K. Lee, N. McEvoy and G. S. Duesberg, *Solid State Electron.*, 2016, **125**, 39-51.
53. S. Lin, Y. Liu, Z. Hu, W. Lu, C. H. Mak, L. Zeng, J. Zhao, Y. Li, F. Yan, Y. H. Tsang, X. Zhang and S. P. Lau, *Nano Energy*, 2017, **42**, 26-33.
54. M. O'Brien, N. McEvoy, T. Hallam, H.-Y. Kim, N. C. Berner, D. Hanlon, K. Lee, J. N. Coleman and G. S. Duesberg, *Sci. Rep.*, 2014, **4**, 7374.
55. C. Yim, M. O'Brien, N. McEvoy, S. Riazimehr, H. Schäfer-Eberwein, A. Bablich, R. Pawar, G. Iannaccone, C. Downing, G. Fiori, M. C. Lemme and G. S. Duesberg, *Sci. Rep.*, 2014, **4**, 5458.
56. G. Mirabelli, L. A. Walsh, F. Gity, S. Bhattacharjee, C. P. Cullen, C. Ó Coileáin, S. Monaghan, N. McEvoy, R. Nagle, P. K. Hurley and R. Duffy, *ACS Omega*, 2019, **4**, 17487-17493.
57. A. K.-V. S. W. G. Alexander V. Naumkin and J. P. Cedric, *NIST X-ray Photoelectron Spectroscopy Database*, 2000, DOI: 10.18434/T4T88K.
58. M. Y. Smirnov, A. V. Kalinkin, A. V. Pashis, I. P. Prosvirin and V. I. Bukhtiyarov, *J. Phys. Chem. C*, 2014, **118**, 22120-22135.
59. J. C. Summers, *Environ. Sci. Technol.*, 1979, **13**, 321-325.
60. R. Addou, L. Colombo and R. M. Wallace, *ACS Appl. Mater. Interfaces*, 2015, **7**, 11921-11929.
61. I. P. Koutsaroff, M. Zelner, P. Woo, L. McNeil, M. Buchbinder and A. Cervin-Lawry, *Integr. Ferroelectr.*, 2002, **45**, 97-103.
62. N. Abe, Y. Otani, M. Miyake, M. Kurita, H. Takeda, S. Okamura and T. Shiosaki, *Jpn. J. Appl. Phys.*, 2003, **42**, 2791.
63. C. Yim, V. Passi, M. C. Lemme, G. S. Duesberg, C. Ó Coileáin, E. Pallecchi, D. Fadil and N. McEvoy, *npj 2D Materials and Applications*, 2018, **2**, 5.
64. G. H. Ryu, J. Chen, Y. Wen and J. H. Warner, *Chem. Mater.*, 2019, **31**, 9895-9903.

**Supplementary Information:**

**Synthesis and characterisation of thin-film platinum disulfide and platinum sulfide**

*Conor P. Cullen[1,2], Cormac Ó Coileáin [1,2], John B. McManus[1,2], Oliver Hartwig[3], David McCloskey[4], Georg S. Duesberg[1,2,3], Niall McEvoy[1,2]*

[1] School of Chemistry, Trinity College Dublin, Dublin 2, D02 PN40, Ireland

[2] AMBER Centre, CRANN Institute, Trinity College Dublin, Dublin 2, Ireland

[3] Institute of Physics, EIT 2, Faculty of Electrical Engineering and Information Technology, Universität der Bundeswehr München, 85579 Neubiberg, Germany

[4] School of Physics, Trinity College Dublin, Dublin 2, Ireland

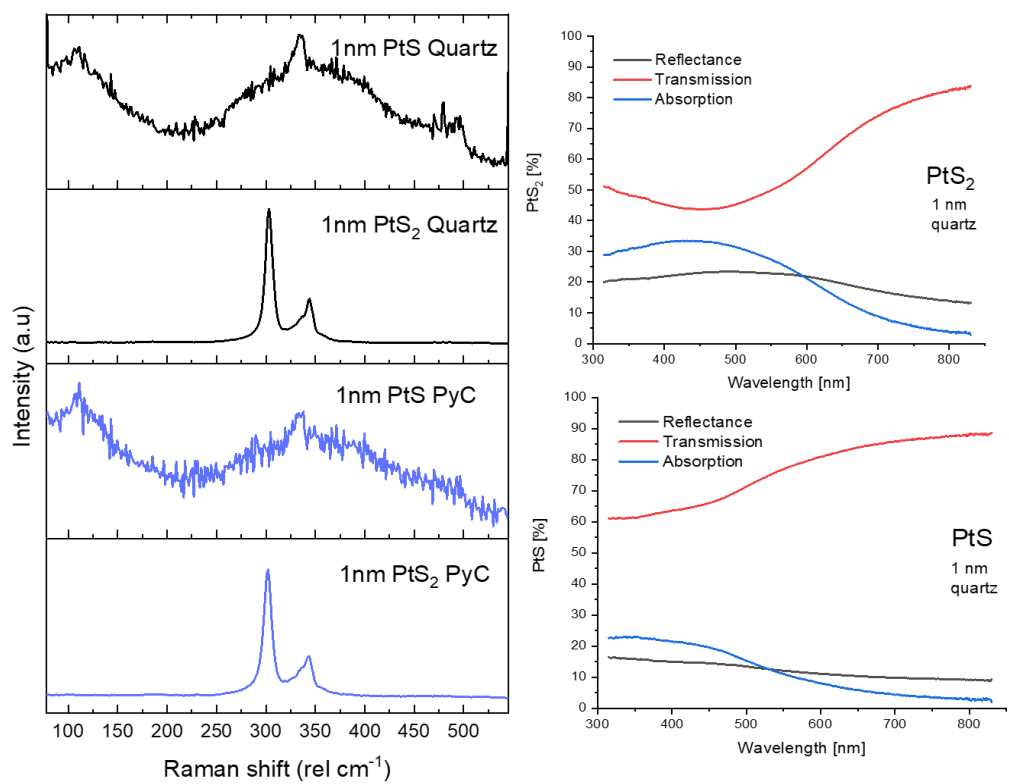

**Figure S1. (a)** Raman spectra of 1 nm PtS$_2$ and PtS films synthesized on quartz and PyC substrates acquired with 532 nm excitation. **(b)** Optical spectra measured for platinum sulfide films synthesized on quartz.

Fig.S1(a) shows Raman spectra which indicate the successful synthesis of PtS and PtS$_2$ on quartz and pyrolytic carbon (PyC) substrates.

1 nm PtS$_2$ and PtS films were deposited on quartz substrates to investigate their optical properties. Fig.S1(b) shows reflectance, transmission and absorption spectra for both films between 314-830 nm. The PtS$_2$ film shows relatively uniform reflectivity across the range at ~20%. An absorption maximum is seen of 34% at ~440 nm while being >70% transparent for wavelengths greater than ~680 nm. The PtS film has lower, but also roughly uniform, reflectivity of ~15%. PtS also shows consistently lower absorption across the range with a maximum of 23% at ~350 nm and >70% transmission for wavelengths >490 nm. The measurements were taken using a UV-vis spectrophotometer with an integrating sphere attachment (Perkin Elmer LAMBDA 650). A lens system was used to collect the spectrum of a 1 mm diameter area. The total diffuse transmission and specular and diffuse reflection were measured. The difference between these two measurements was used to calculate the absorption of the films.

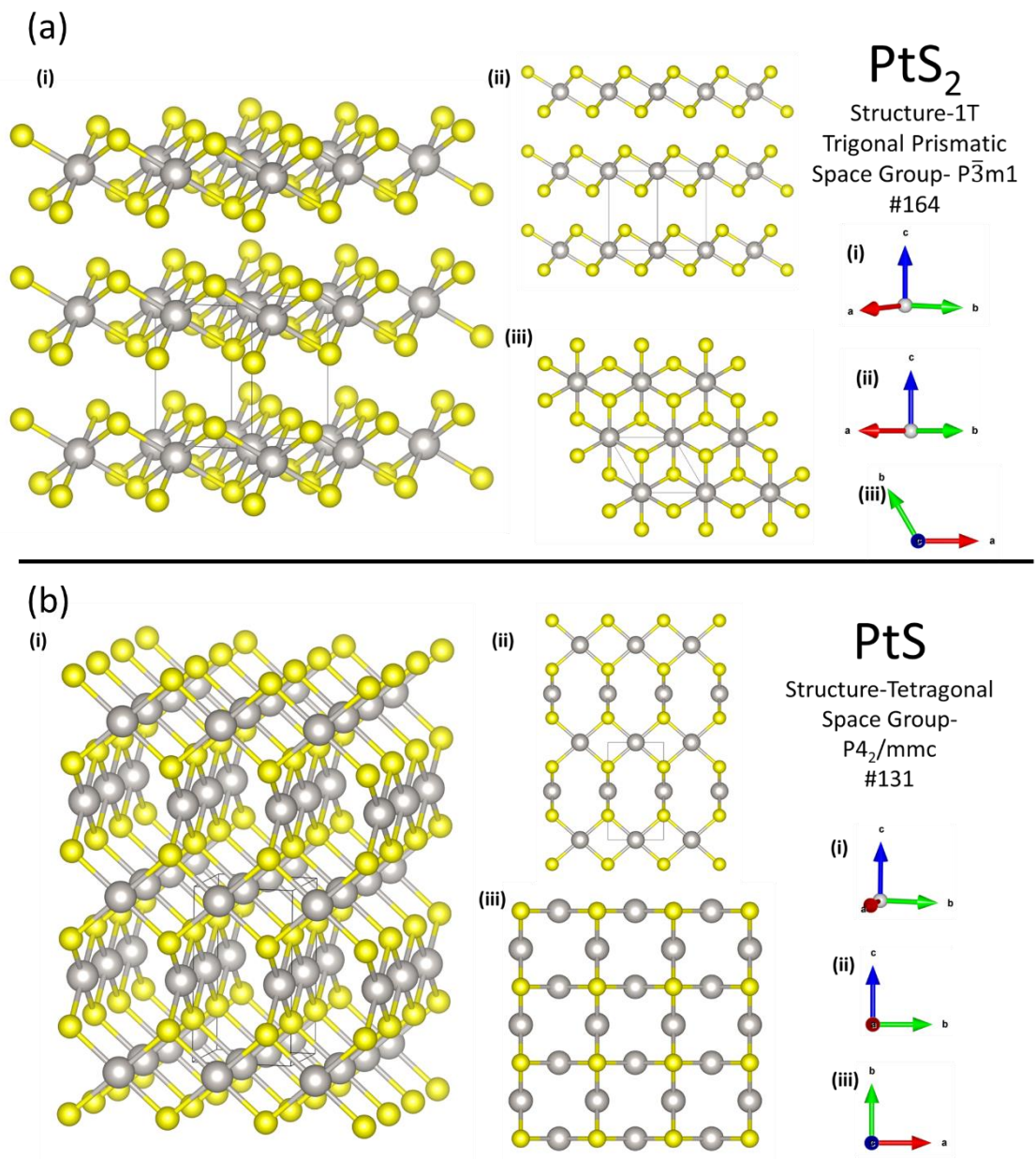

**Figure S2.** Diagrams of the atomic structure for **(a)** PtS$_2$ and **(b)** PtS

**Structural representations**

Atomic structural representations for each material were generated using VESTA 3 software.[1]

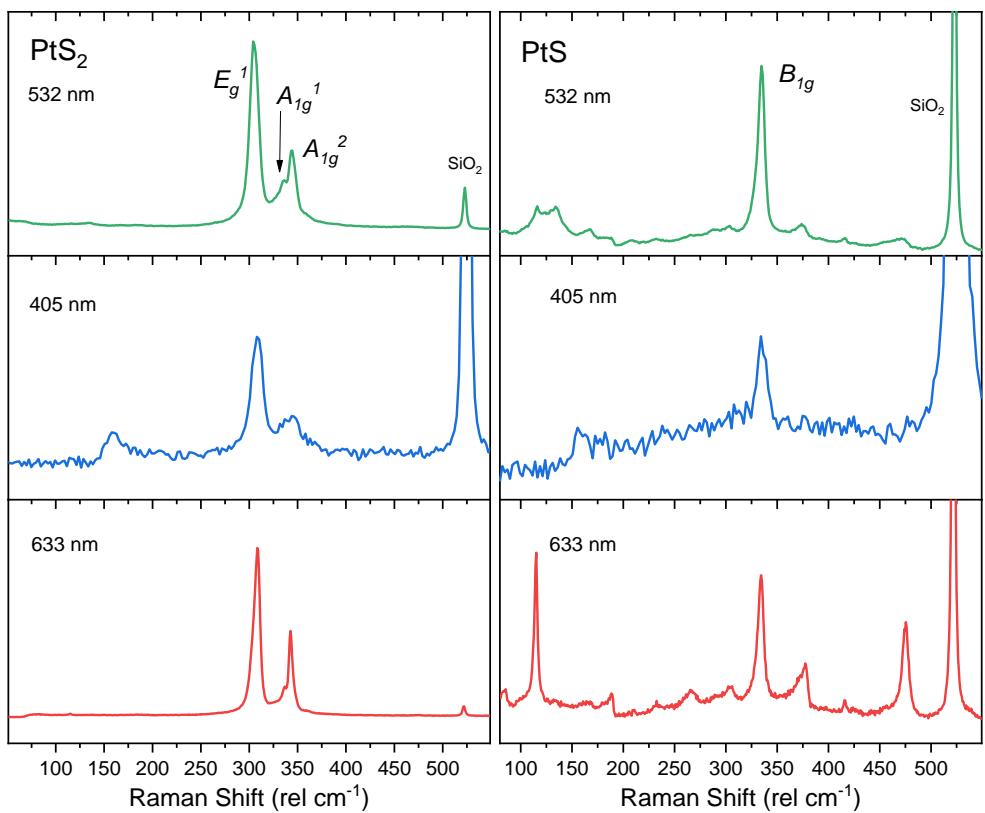

**Figure S3. (a)** Raman spectra of a 5 nm PtS$_2$ film using different wavelength Raman lasers. **(b)** Raman spectra of a 5 nm PtS film using different wavelength Raman lasers

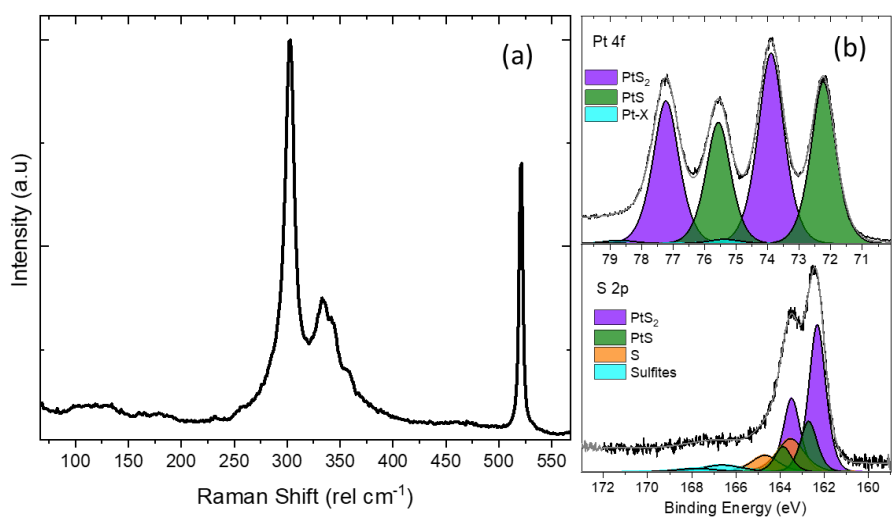

**Figure S4. (a)** Raman spectra of a mixed PtS$_2$-PtS film grown at 400 °C. **(b)** XPS Pt 4f and S 2p core-level spectra of the mixed film showing the relevant components of each material

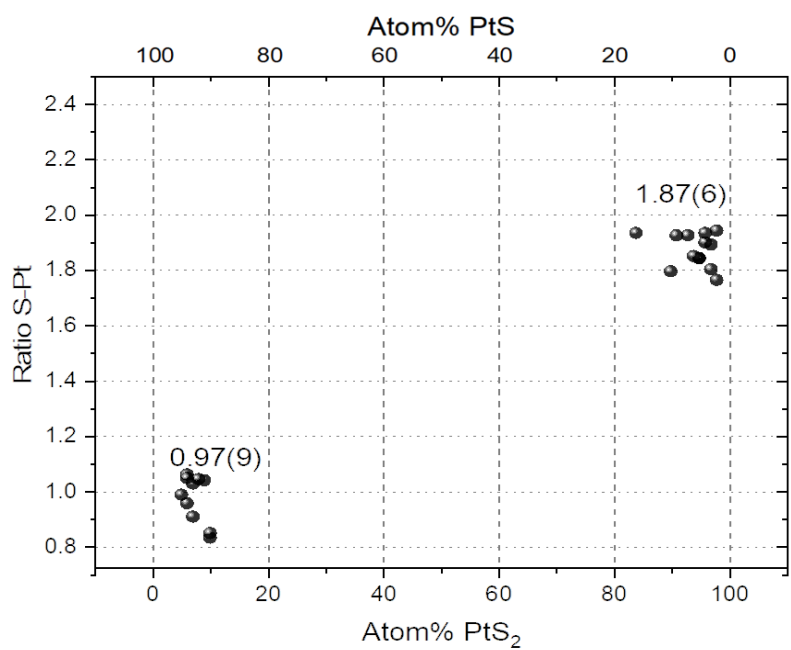

**Figure S5.** Plot of the S-Pt ratio for 24 Pt sulfide films as calculated by comparison of relative XPS peak areas showing the clusters for PtS and PtS$_2$.

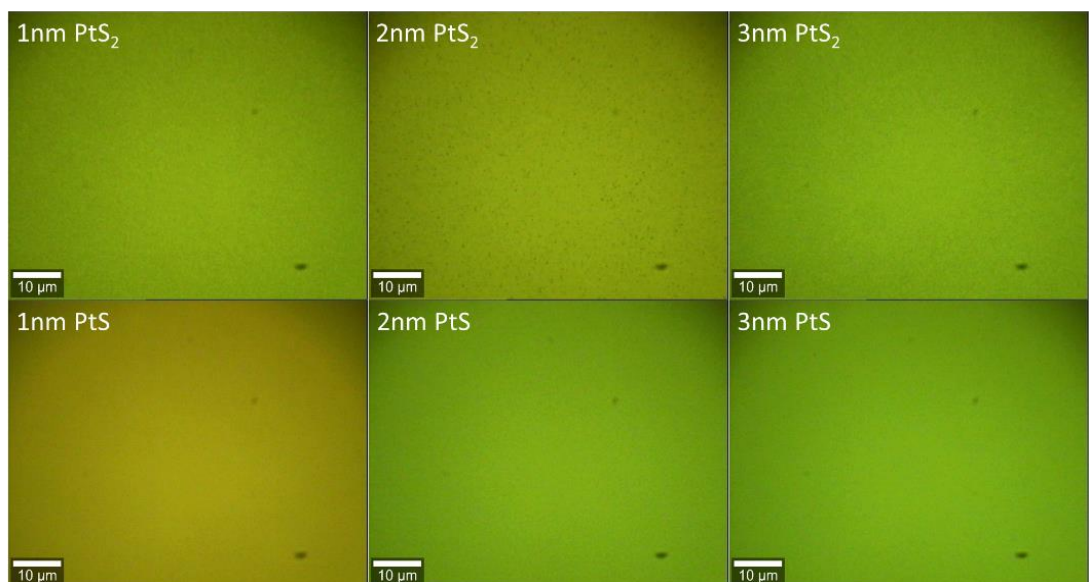

**Figure S6.** Optical images with a 100x microscope objective of the surface of 1, 2, and 3 nm PtS$_2$ and PtS films.

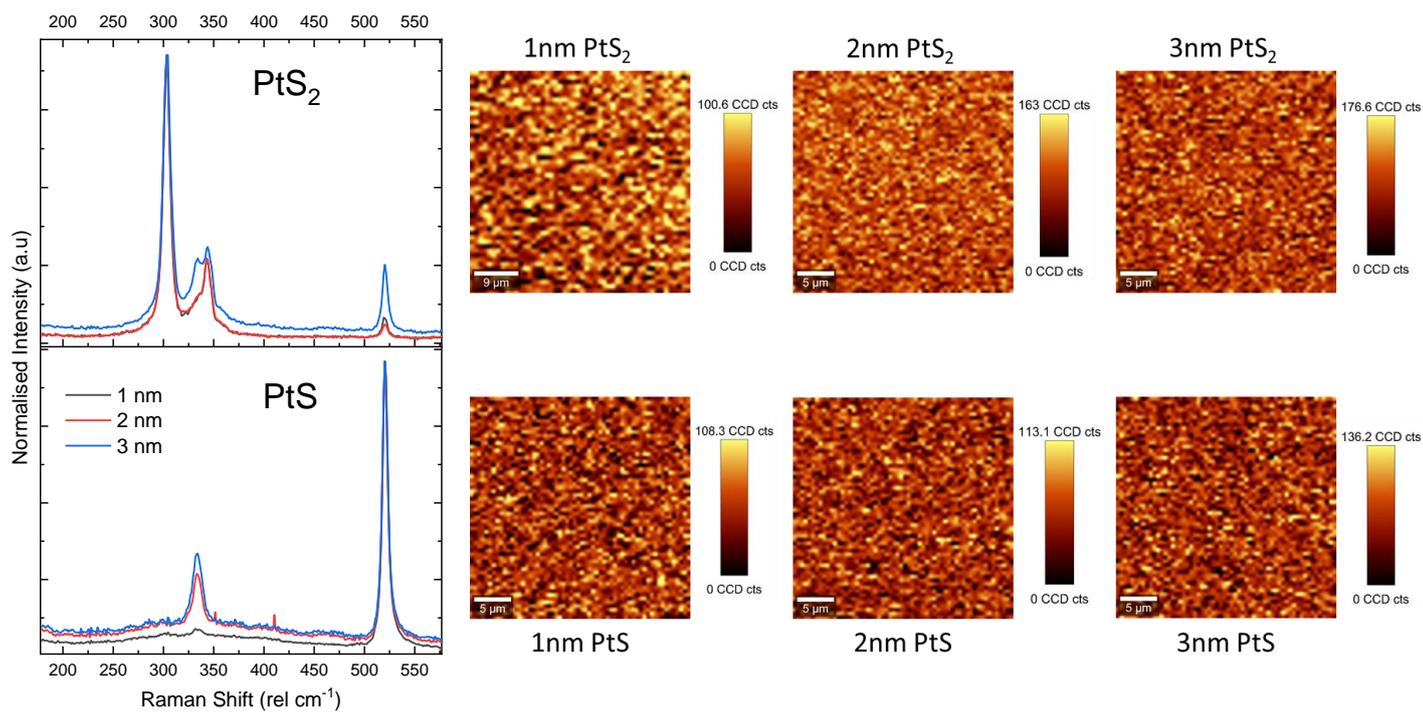

**Figure S7.** Averaged Raman spectra for 1, 2, and 3 nm $PtS_2$ and PtS films. Raman intensity maps of the surface of the films show consistent signal across the surface for all films.

## 2nm PtS$_2$ Raman data

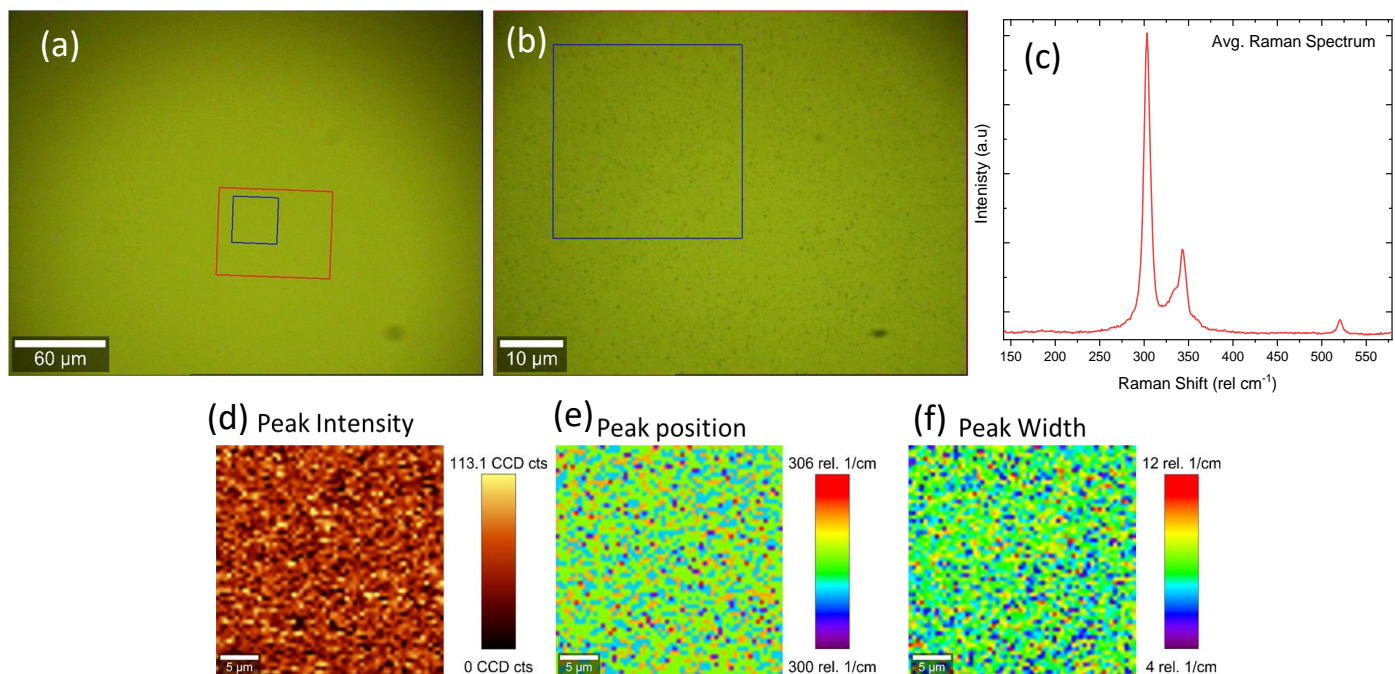

**Figure S8. (a)** Optical image with a 20x microscope objective of the surface of a 2 nm PtS$_2$ film, with a red box showing the area in **(b)** and a blue box showing the area mapped by Raman spectroscopy. **(b)** A 100x microscope objective image, with the blue box representing the area mapped by Raman spectroscopy. **(c)** Average Raman spectra of the film. **(d)** $E_g^1$ Raman peak intensity map. **(e)** Raman peak position map. **(f)** Raman peak width (FWHM) map.

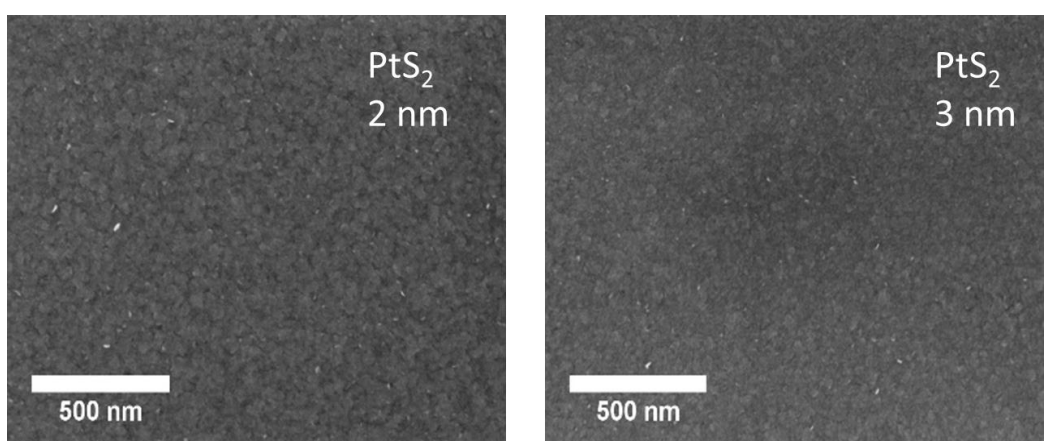

**Figure S9.** SEM images of PtS$_2$ films synthesized from 2 and 3 nm Pt films.

## AFM analysis of Pt, PtS and PtS$_2$ films

Atomic force microscopy (AFM) was performed on a Bruker Multimode 8 with ScanAsyst Air AFM probes in ScanAsyst Air 146 mode. The applied scan rate was 1 Hz with an image resolution of 512x512 points. Peak-force tapping mode was used to obtain the topography. The acquired data was processed using Gwyddion software.[2]

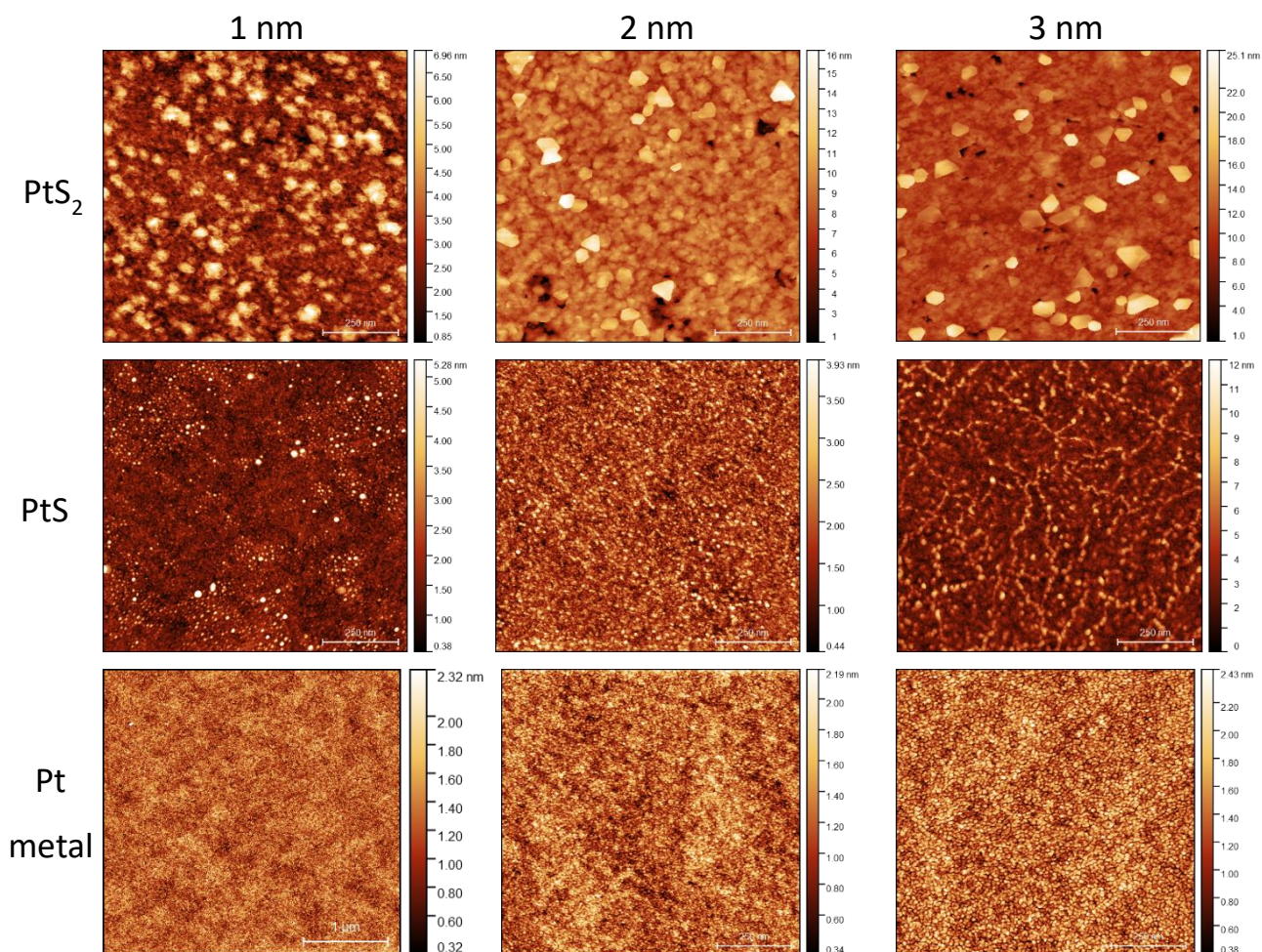

**Figure S10.** AFM maps of the surface of 1, 2, and 3 nm PtS$_2$, PtS, and as-deposited Pt metal films. Scale bar 250 nm, except for 1 nm Pt metal (1 μm).

Determining the theoretical film thickness after conversion of a Pt film to PtS$_2$ or PtS is difficult due to the random orientation of the crystallites and the layered/non-layered nature of the materials. To roughly estimate the film expansion, we use the unit cell volume for the respective materials, determined using literature values for PtS$_2$,[3] PtS,[4] and Pt metal.[5] This yields an estimate for film volume expansion of 3.5x for PtS$_2$ and 4.7x for a PtS film. This greater thickness for PtS films over PtS$_2$ is also seen in our measured film thicknesses. PtSe$_2$ synthesised through a similar synthesis process was found to have an expansion of ~4x from the as-deposited platinum.[6]

**Table S1.** AFM measured film properties.

| Film | 1nm Pt metal | 1nm PtS$_2$ | 1nm PtS | 2nm Pt metal | 2nm PtS$_2$ | 2nm PtS | 3nm Pt metal | 3nm PtS$_2$ | 3nm PtS |
|---|---|---|---|---|---|---|---|---|---|
| Roughness (nm) | 0.297 | 1.04 | 0.63 | 0.3 | 1.57 | 0.52 | 0.35 | 2.45 | 1.33 |
| Expansion factor | - | 1.4 | 3.1 | - | 3.2 | 5.9 | - | 2.5 | - |
| Thickness (nm) | 1.13 | 1.57 | 3.53 | 2.43 | 7.77 | 14.37 | 3.74 | 9.37 | - |

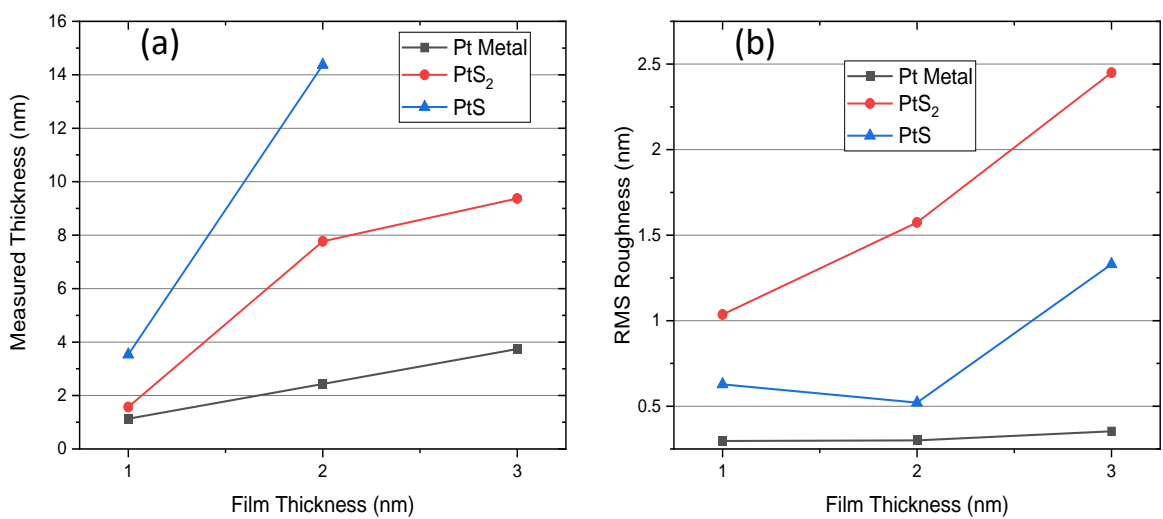

**Figure S11. (a)** Graph of measured film thickness and **(b)** surface RMS roughness against intended initial Pt metal thickness for the AFM images in Fig.S10.

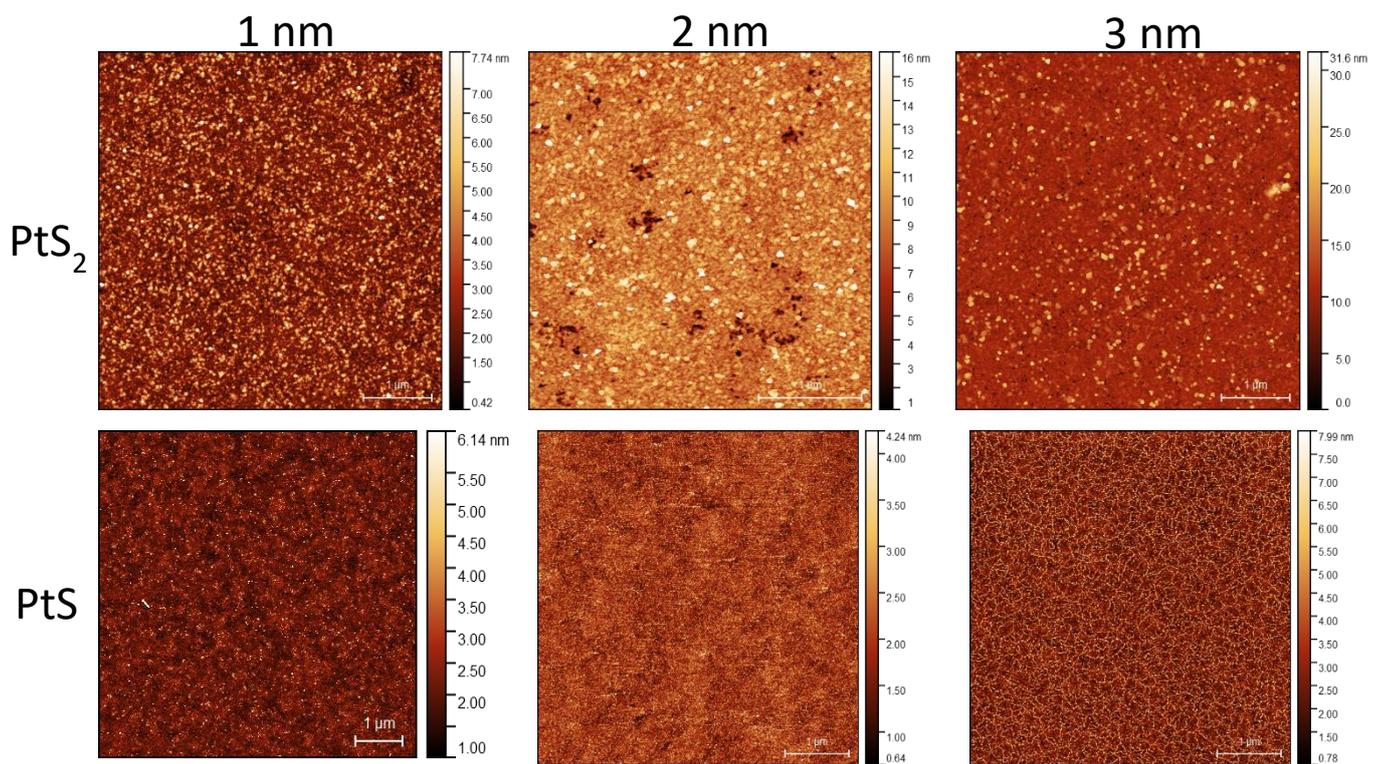

**Figure S12.** AFM maps of larger areas of the surface of 1, 2, and 3 nm PtS$_2$ and PtS films. Scale bar 1 μm.

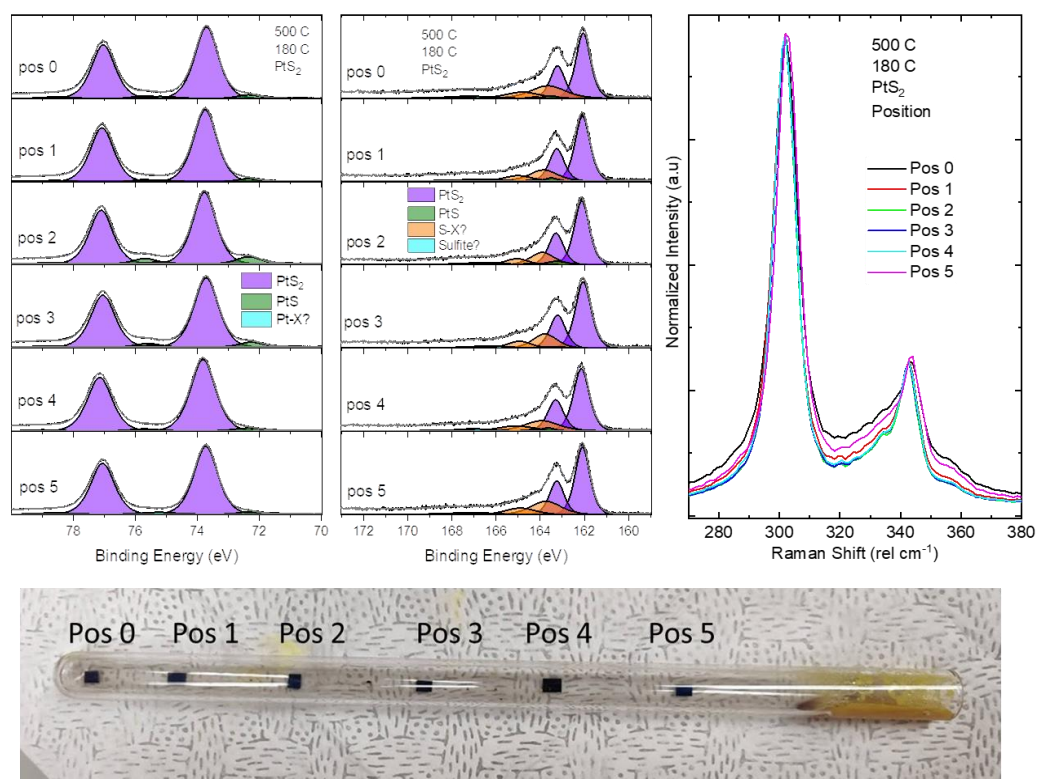

**Figure S13. (a)** XPS spectra of 1 nm PtS$_2$ films synthesized along the length of the internal quartz tube. **(b)** Raman spectra of the PtS$_2$ films. **(c)** Picture of the quartz tube showing the positions of the substrates.

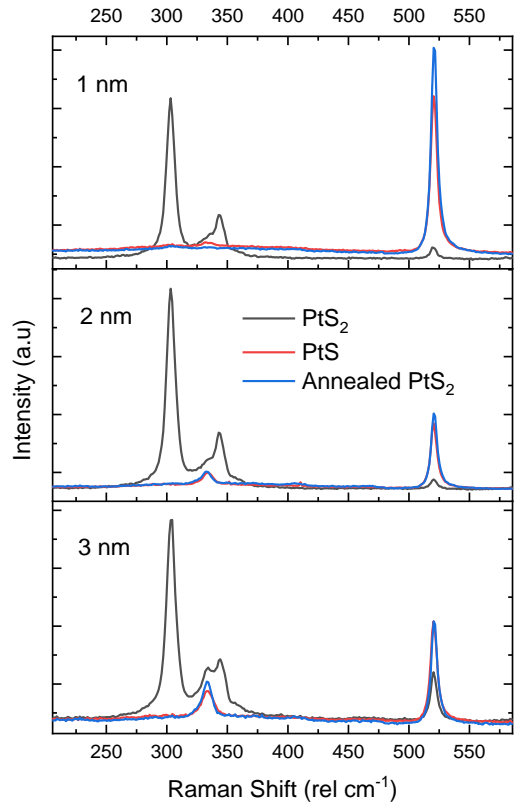

**Figure S14.** Raman spectra for 1, 2, and 3 nm films of $PtS_2$, PtS, and $PtS_2$ films after 600 °C annealing in an inert environment at ~1 mbar for 30 minutes, showing a change from $PtS_2$ to PtS Raman signal.

**References**

S1.  K. Momma and F. Izumi, *J. Appl. Crystallogr.*, 2011, **44**, 1272-1276.

S2.  D. Nečas and P. Klapetek, *Cent. Eur. J. Phys.*, 2012, **10**, 181-188.

S3.  H. A. H. Mohammed, G. M. Dongho-Nguimdo and D. P. Joubert, *Materials Today Communications*, 2019, DOI: 10.1016/j.mtcomm.2019.100661, 100661.

S4.  V. I. Rozhdestvina, A. V. Ivanov, M. A. Zaremba, O. N. Antsutkin and W. Forsling, *Crystallogr. Rep.*, 2008, **53**, 391-397.

S5.  W. P. Davey, *Phys. Rev.*, 1925, **25**, 753-761.

S6.  C. Yim, V. Passi, M. C. Lemme, G. S. Duesberg, C. Ó Coileáin, E. Pallecchi, D. Fadil and N. McEvoy, *npj 2D Materials and Applications*, 2018, **2**, 5.